\documentclass[namedreferences]{SolarPhysics}
\usepackage[optionalrh]{spr-sola-addons} 
\usepackage{graphicx}        
\usepackage{color}           
\usepackage{url}             

\newcommand{\etal}{{\it et al.}}
\newcommand{\ie}{{\it i.e.}}
\newcommand{\cf}{{\it cf.}}
\newcommand{\exg}{{\it e.g.}}

\newcommand{\aap}{    {\it Astron. Astrophys.}}

\newcommand{\apj}{    {\it Astrophys. J.}}
\newcommand{\apjs}{   {\it Astrophys. J. Suppl.}} 

\newcommand{\araa}{   {\it Annu. Rev. Astron. Astrophys.}} 
\newcommand{\basi}{   {\it Bul. Astron. Soc. India}}

\newcommand{\jaa}{    {\it J. Astrophys. Astron.}} 

\newcommand{\jgr}{    {\it J. Geophys. Res. (Space Physics)}}
\newcommand{\mnras}{  {\it Mon. Not. Roy. Astron. Soc.}}
\newcommand{\nat}{    {\it Nature}}

\newcommand{\pasj}{   {\it Publ. Astron. Soc. Japan}}

\newcommand{\solphys}{{\it Solar Phys.}}

\newcommand{\SunGeosph}{{\it Sun Geosph.}}

\begin{document}

\begin{article}  

\begin{opening}

\title{Transient Magnetic and Doppler Features Related to the White-light Flares in NOAA 10486}

\author{R. A.~\surname{Maurya}$^{1}$\sep
        A. ~\surname{Ambastha}$^{2}$\sep
         }
\runningauthor{Maurya and Ambastha}
\runningtitle{Transient Magnetic and Doppler Features}

\institute{$^{1}$Udaipur Solar Observatory (Physical Research Laboratory),\\ P.O. Box 198, Dewali, Badi Road, Udaipur 313 001, INDIA.\\
              $^{1}$ e.mail: \url{ramajor@prl.res.in}\\
              $^{2}$ e.mail: \url{ambastha@prl.res.in}\\
             }

\begin{abstract}
Rapidly moving transient features have been detected in magnetic and Doppler images of super-active region NOAA 10486 during the X17/4B flare of 28 October 2003 and the X10/2B flare of 29 October 2003. Both these flares were extremely energetic white-light events. The transient features appeared during impulsive phases of the flares and moved with speeds ranging from 30 to 50 km s$^{-1}$. These features were located near the previously reported compact acoustic \cite{Donea05} and seismic sources \cite{Zharkova07}. We examine the origin of these features and their relationship with various aspects of the flares, {\it viz.}, hard X-ray emission sources and flare kernels observed at different layers - (i) photosphere (white-light continuum), (ii) chromosphere (H$\alpha$ 6563\AA), (iii) temperature minimum region (UV 1600\AA), and (iv) transition region (UV 284\AA).  
\end{abstract}
\end{opening}
\section{Introduction}
     \label{S-Introduction}
Catastrophic changes in the coronal magnetic field topology are believed to trigger energetic transient events, such as, solar flares and CMEs. The MHD catastrophe leads to the reconnection of magnetic field lines in the corona that results in a wealth of observed post-flare phenomena. As flares derive their energy from stressed magnetic field, it is expected that the magnetic stress or non-potentiality would relax toward a lower energy state after the release of excess energy available in the active region. It was first suggested several decades ago by \inlinecite{Giovanelli39} that flare energy release should be associated with observable magnetic field changes. Extensive efforts have been made since then to detect such changes using photospheric magnetic field measured by ground-based instruments  \cite{Patterson81,Patterson84,Wang92,Ambastha93,Chen94,Hagyard99}, and more recently, by space-borne instruments \cite{Kosovichev01,Qiu03,Wang06}.
\par A wide variety of results have been reported on the pre- and post-flare changes in magnetic field parameters. However, these remained mostly unreliable because of poor sensitivity, spatial resolution, cadence and coverage of the available magnetographs. Furthermore, serious questions have also been raised about the nature of observed magnetic field changes as these measurements are expected to be affected by flare-induced line profile changes \cite{Harvey86,Qiu03}. Nevertheless, with the availability of recent high-quality, high-cadence magnetic field observations, there is a mounting evidence of rapid, permanent changes in longitudinal and transverse magnetic fields during the course of large flares \cite{Sudol05}. 
\par There are some other physical processes accompanying the sudden energy release occurring in the solar corona that may influence the magnetic field measurements. A large number of charged particles, \ie, electrons and protons, accelerated during the flare energy release in the corona can precipitate along magnetic field-lines and lose their energy in the lower atmosphere. During this process, microwave radiation is produced by gyrosynchrotron process and hard X-ray (HXR) emission through a thick-target bremsstrahlung process \cite{Brown71,Emslie78}. White-light flares (WLFs), the most energetic of all flares, were suggested to be associated with direct heating by nonthermal or accelerated particles, specifically quasi-relativistic electrons \cite{Rust75,Hudson92,Neidig93}. This is observationally supported by the HXR footpoints, \ie, sites of nonthermal particle acceleration, matching well with WLF ribbons  \cite{Fletcher01,Metcalf03}. However, only very high energy electrons having energy exceeding a few MeV are expected to penetrate to deeper photospheric layers due to the increasing density. Only these high energy electrons can contribute to direct heating that may not be adequate to produce WLF emission. It was inferred from the TRACE\footnote{Transition Region and Coronal Explorer} WL and {\it Yohkoh}/HXT data that WLF ribbons originate in the chromosphere and the temperature minimum region, and that the enhanced WL emission is caused by ionization and subsequent recombination of hydrogen. Energy thus deposited in the chromosphere by the electron beam is then transported to the lower atmosphere by a back-warming process. Observed changes in magnetic field parameters during the impulsive phase of a flare would be affected by a variety of these effects.

\par Localized sign reversal of magnetic polarity, termed as  ``magnetic anomaly'', has been observed during the impulsive phase of some flares \cite{Qiu02,Qiu03}. Such a sign reversal has also been attributed to the change in the spectral line profile, from absorption to emission, by numerical models \cite{Machado80,Vernazza81,Ding89,Ding02}. However, non-LTE calculations for the spectral line Ni~{\sc i} 6768\AA, used for magnetic field measurements in GONG\footnote{Global Oscillation Network Group} and SOHO/MDI\footnote{Solar and Heliospheric Observatory, Michelson Doppler Imager} instruments, have shown that this absorption line can turn into emission only by a large increase of electron density and not by heating of the atmosphere by any other means.  Consistent with this, \inlinecite{Qiu03} discovered sign reversals in locations of strong HXR emission formed near the cooler sunspot umbral/penumbral areas. Injection and transport of high-energy electrons was observationally inferred from a M9.8 flare observed in microwave by Owens Valley Solar Array, and in HXR by hard X-ray telescope on board {\it Yohkoh} \cite{Lee02}. Edelman \etal~\shortcite{Edelman04} simulated the GONG and MDI observations for solar flares and concluded that as compared to Doppler velocity measurements the magnetic field measurements are less sensitive to the line shape changes. In addition, their numerical simulation also showed that the observed transient sign reversal may be produced when the Ni~{\sc i} 6768\AA~line profile turns into emission as a result of non-thermal beam impact on the atmosphere in regions of strong magnetic fields. The effect of line profile changes on magnetic field estimates has been reported also by \inlinecite{Abramenko04}. 

\par More recently, flare-associated helioseismic effects have been reported using different local helioseismology techniques, \exg, detection of acoustic sources  by holography \cite{Donea05}, flare-induced seismic waves using time-distance analysis \cite{Kosovichev06} and post-flare amplitude enhancement of acoustic or p-modes by ring-diagram technique \cite{Ambastha03,Ambastha04}.  \inlinecite{Zharkova07} have found seismic sources to be located near the acoustic sources discovered earlier by \inlinecite{Donea05} during the large flare in NOAA 10486. Several other flare-productive regions have also been extensively studied.

\par While examining the GONG magnetogram movie of the X17/4B flare of 28 October 2003 in NOAA 10486, \sloppy rapidly ``moving'' transient magnetic  features (MFs) were detected during the impulsive phase \cite{Ambastha07a,Maurya08}. These interesting transient features moved away from the flare site with separation velocities ranging from 30 to 50 km s$^{-1}$ comparable to the velocity of seismic waves reported by \inlinecite{Kosovichev98}. These MFs do not appear to be instrumental artifacts as GONG and MDI movies both exhibited these features. The exact nature of the MFs is, however, not well understood and further investigations are required. If the MFs were indeed associated with the white-light flare by way of the absorption line profile turning into emission, they are expected to be correlated with WLF kernels. In particular, it would also be interesting to examine whether these features were essentially ``moving'' sign-reversal anomalies associated with HXR sources produced by nonthermal electron beams \cite{Qiu02}. It is to note that we also detected ``Doppler velocity transients'' during the impulsive phase of this flare from GONG and MDI Dopplergram movies. These are similar to the localized Doppler velocity enhancements, co-spatial with H$\alpha$ flare kernels, earlier reported by \inlinecite{Venkatakrishnan08}.
\par	In this paper, we discuss spatial and temporal correlations of the ``moving'' magnetic and Doppler velocity transient features observed during the energetic white-light flares of 28 and 29 October 2003.  We examine the properties of these MFs in the light of recent studies related to flare-associated changes in magnetic and Doppler velocity fields. We also attempt to find their  correlation with flare kernels as observed in the lower atmosphere, \ie,  photosphere and chromosphere and the corresponding features in upper layers using the available multi-wavelength observations. In Section~\ref{S-Data}, we describe the observational data used in this study. The data analysis technique  is given in Section~\ref{S-analysis}. Properties of MFs and associated flare kernels are described in Sections~\ref{S-flare28}--\ref{S-flare29}. Finally, discussions and conclusions are presented in Section~\ref{S-Discussions}.
\section{The Observational Data}
   \label{S-Data}
We have used magnetograms, Dopplergrams and white-light images obtained by GONG and SOHO/MDI to study the WL flares of 28 and 29 October 2003. These observations provide information on the temporal and spatial evolution of photospheric magnetic field, Doppler velocity field and intensity in NOAA 10486. To examine properties of the magnetic and Doppler MFs and their correlation with the flare kernels, we have used white-light intensity images for the photospheric layer. For upper atmospheric layers, we have used TRACE UV 1600\AA~for the temperature minimum region, chromospheric H$\alpha$ observations obtained from USO\footnote{Udaipur Solar Observatory} and MLSO\footnote{Mauna Loa Solar Observatory},  and TRACE UV 284\AA~for the transition region. To compare the locations of flare kernels and MFs with HXR sources, we have used HXR data obtained by RHESSI\footnote{Reuven Ramaty High-Energy Solar Spectroscopic Imager}. Details of these observational data are described below: 
\begin{itemize}
\item {\bf GONG data:}~Magnetograms, Dopplergrams and white-light images obtained from GONG instrument have spatial and temporal resolutions of 2.5 arcsec/pixel and 1 min, respectively \cite{Harvey88}. The instrument uses spectral line centered at Ni~{\sc i} 6768\AA~and works on the principle of phase shift interferometry (Harvey 2008, private communication).	
\item {\bf SOHO/MDI data:}~In addition to the GONG data, we have also used the corresponding SOHO/MDI data \cite{Scherrer95} to inter-compare the observed changes in magnetic, Doppler velocity and intensity maps. The SOHO/MDI data have spatial resolution of 2.0 arcsec/pixel and temporal cadence of 1 min. 
 \item{\bf USO H$\alpha$:}~We have used high spatial and temporal resolution H$\alpha$ filtergrams obtained from USO, Udaipur (India) for studying the chromospheric evolution of the X17/4B flare of 28 October 2003. These filtergrams were taken by a 15-cm aperture f/15 Spar telescope at a cadence of 30 s during the quiet phase and 3 s in flare mode. The spatial resolution of USO filtergrams is 0.4 arcsec/pixel at the CCD detector plane. 
\item{\bf MLSO H$\alpha$:}~MLSO H$\alpha$ filtergrams have been used for the X10/2B flare of 29 October 2003/20:49 UT as this event occurred during the night-time of the USO, Udaipur site. This data has spatial and temporal resolutions of 2.9 arcsec/pixel and 3 min, respectively.
 \item{\bf TRACE UV:}~The data corresponding to the temperature minimum and transition regions were obtained from TRACE \cite{Handy99,Schrijver99} which correspond to 1600\AA~and 284\AA~data sets, respectively. These images have spatial and temporal resolutions of the order of 0.5 arcsec/pixel and 35 s, respectively.
\item{\bf RHESSI HXR:}~We have used X-ray data obtained from RHESSI \cite{Lin02} which observes solar HXR and $\gamma$-rays from 3 keV to 17 MeV with spatial resolution as high as $\approx$ 2.3 arcsec with a full-Sun field of view. The raw format data (\ie, modulated Fourier components) for both events were obtained from the RHESSI data site. HXR maps for the events were then constructed using Pixon method \cite{Hurford02} under Solar Software (SSW) for the dimension of $64\times64$ pixels centered at the flaring regions. The spatial and temporal resolutions of the constructed images are 4 arcsec and 1 min, respectively. We chose different energy bands for the two flares due to their differing energetics: higher energy (100-200 keV) for the 28 October 2003 event and the lower energy (50-100 keV) for the 29 October 2003 event. 
 
\end{itemize}
\section{Data Analysis}
    \label{S-analysis}

We have used time-series of full-disk GONG and MDI magnetograms and Dopplergrams for identifying various transient features during the course of evolution of the energetic flares of 28 and 29 October 2003~in NOAA 10486. Full disk images were aligned by de-rotating with the solar rotation. We ignored the effect of solar differential rotation while correcting the images, since (i) the active region was located close to the solar equator, (ii) selected areas-of-interest were much smaller than the disk, and (iii) durations of the peak phase of flare-events were relatively short, \ie, 10-15 min. 

\begin{figure}[ht] 
	\centering
\includegraphics[width=1.0\textwidth]{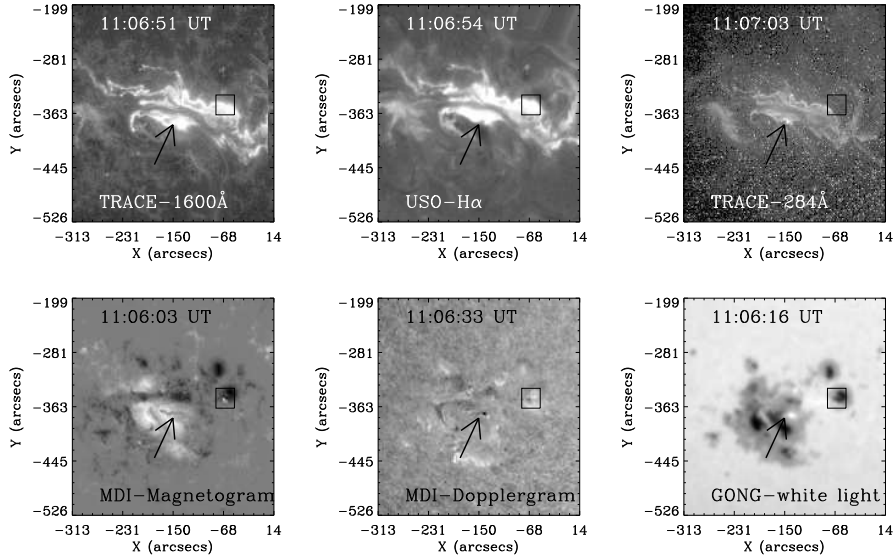}
	\caption{TRACE UV 1600\AA, USO H$\alpha$ and TRACE UV 284\AA~(top row, from left to right columns), and the MDI magnetogram, Dopplergram and GONG white-light image (bottom row, from left to right columns) obtained around the peak phase of the white-light X17/4B flare of 28 October 2003. The arrow and the box indicate locations of observed ``moving'' magnetic transients.}
	\label{fig:immos28}
\end{figure}

We applied a cross-correlation method based on the Fourier technique to obtain co-aligned magnetograms registered within a sub-pixel accuracy. All the images were aligned using this procedure. We overlaid contours of magnetic flux on various observational images for identifying and selecting corresponding areas-of-interest by visual inspection of various structures, such as, filaments and sunspots. Figures~\ref{fig:immos28} and \ref{fig:immos29} show the selected areas of NOAA 10486 from TRACE UV 1600\AA, USO H$\alpha$, TRACE UV 284\AA, MDI magnetogram/Dopplergram, and GONG white-light images for the flares of 28 and 29 October 2003, respectively. 
\section{The X17/4B Flare of 28 October 2003/11:10 UT}
  \label{S-flare28}
 The two-ribbon X17/4B flare of 28 October 2003 was one of the most energetic events that occurred in NOAA 10486. The active region was located  close to the solar disk center at the time of this flare; therefore, projection effects did not pose any serious difficulty in the analysis. According to GOES\footnote{Geostationary Operational Environmental Satellite} X-ray flux data, this flare started at 09:51 UT and ended at 11:24 UT, with the maximum phase at 11:10 UT. However, as seen in H$\alpha$, the flare lasted over a considerably longer period.  
 
\par Significant changes were observed in photospheric magnetic and Doppler velocity fields as seen from the movies of GONG and MDI observations for 28 October 2003. These included gradual, evolutionary changes in the active region during the course of the day. In addition, there were also abrupt and persistent changes that occurred at some localized sites in the neighborhood of flare kernels observed in white-light \cite{Ambastha07a}. Figure~\ref{fig:immos28} shows multi-wavelength images corresponding to different layers of solar atmosphere obtained during the peak phase of this flare. The top row, left to right columns, shows the flare kernels observed in the temperature minimum region, the chromosphere and the transition region. It may be noted that the flare kernels displayed structural similarity in these layers. Images corresponding to the photospheric level are shown in the bottom row, left to right columns: MDI magnetogram, MDI Dopplergram and GONG white-light. The white-light image clearly shows emission kernels of this energetic WLF. 

\par From the movie of GONG magnetograms obtained during this super-flare, interesting ``magnetic'' transient features were observed moving rapidly away from the flare site. Figure~\ref{fig:immos28} shows these features as seen at around 11:06 UT. One of these features was observed to be moving toward the leading sunspot (marked by a rectangular box), while, the other moved perpendicular to the magnetic neutral line (marked by an arrow).  These features appeared and faded in a few minutes' period during the impulsive phase of the flare (\opencite{Ambastha07b}, \citeyear{Ambastha08}; \opencite{Maurya08}). Both GONG and MDI images exhibited these moving features.  Similar, albeit fainter, Doppler velocity transients associated with the WLF event were also detected in the Dopplergrams. 
 
 \begin{figure}[ht] 
	\centering
		\includegraphics[width=0.9\textwidth]{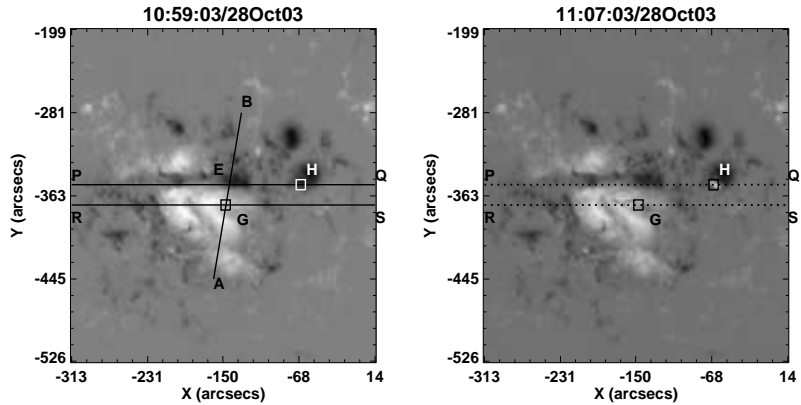}
	\caption{MDI magnetograms taken around the pre- and peak phases of the 28 October 2003 flare at 10:59:03 UT (left panel) and 11:07:03 UT (right panel), respectively. PQ and RS are marked along which magnetic flux profiles are shown in Figure~\ref{fig:fluxbd28}. AB marks the direction of motion of WLF kernels and MFs. Distances are estimated from a reference point E (left panel).}
	\label{fig:imbd28}
\end{figure}
\begin{figure}[ht] 
	\centering
		\includegraphics[width=0.9\textwidth]{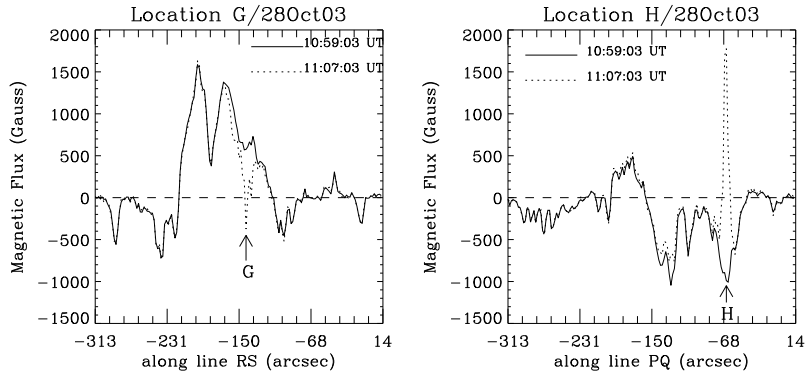}
	\caption{Magnetic flux along lines RS (left panel) and PQ (right panel) shown in Figure~\ref{fig:imbd28}. Solid and dotted profiles represent magnetic flux at 10:59:03 UT and 11:07:03 UT, respectively. Polarity reversals at points G and H are evident at 11:07:03 UT, \ie, around the peak phase of the flare. Else where along the two lines, magnetic flux remained nearly unchanged. }
	\label{fig:fluxbd28}
\end{figure}

\subsection{Magnetic Field Transients and the Polarity Sign Reversal}
 \label{S-inversion28}
  Corresponding to the observed moving transients, we noticed a sign reversal of magnetic polarity (also, Doppler velocity) at some locations during the impulsive phase of the flare. To precisely identify locations of sign reversal, we plotted magnetic flux profiles along a horizontal raster moving from the bottom to the top of selected magnetograms during the pre- and peak-phases of the flare. We found two locations G and H around which magnetic polarity sign reversals occurred (Figure~\ref{fig:imbd28}). We have drawn lines RS and PQ passing through these two points over the magnetograms at 10:59:03 UT (solid lines) and 11:07:03 UT (dotted lines), respectively. The corresponding magnetic flux profiles along these lines are shown in Figure~\ref{fig:fluxbd28}. It is evident that magnetic polarity reversals occurred at G and H as the moving ``magnetic'' features passed over these locations at 11:07:03 UT. Else where, magnetic flux matched fairly well along these lines at the two time instants. 

\par Such sign reversals were already reported several years ago by \inlinecite{Zirin81} for some large flares. This was attributed to the heating of lower atmosphere by the flare such that the core of the absorption line, used for measurement, turned to emission \cite{Patterson84}. More recently, this effect was reported for the large X-class flare of 6 April 2001 using MDI magnetograms \cite{Qiu03}. However, sign reversals observed by MDI instrument are difficult to be explained by the sudden heating of the lower atmosphere. This is because Ni~{\sc i} 6768\AA~line (used by MDI and GONG instruments) is formed in the temperature minimum region and is rather stable against temperature changes \cite{Bruls93}. Interestingly, \inlinecite{Qiu03} noted that sign-reversals occurred near locations of strong magnetic field, cooler sunspot umbrae that are exactly co-aligned with HXR sources. This suggested that the sign reversal anomaly was associated with non-thermal electron beam precipitating to the lower atmosphere near cool sunspots. This inference was further supported by a numerical simulation of MDI measurements incorporating the flare effects. 

\par The transient sign-reversals observed during the impulsive phase of the X17/4B flare in NOAA 10486 appear to be similar in nature as reported by \inlinecite{Qiu03}. However, in this case we found two locations that somewhat differed in their basic nature -- (i) the sign reversal at H was located in strong magnetic field region of the leading, negative polarity umbra; similar to that reported by \inlinecite{Qiu03}, and remained nearly stationary during the entire impulsive phase, (ii) the sign reversal at G first appeared near the neutral line, \ie, a weak field location, and subsequently moved rapidly toward stronger magnetic field region of the following, positive polarity sunspots. 
\par In the following section, we examine these moving features and their association with the flare kernels observed at various solar atmospheric layers. 
\begin{figure}
   \centering
\includegraphics[width=0.3\textwidth, height=0.6\textheight,angle=-90]{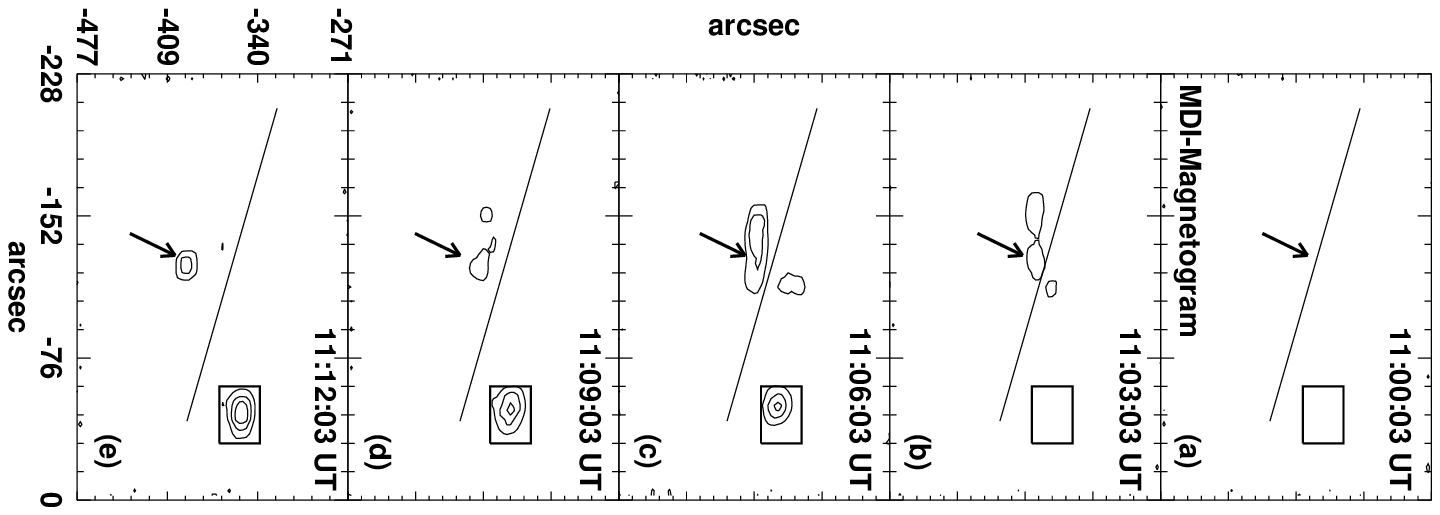}\vspace{-0.061\textheight}\\
\includegraphics[width=0.3\textwidth, height=0.6\textheight,angle=-90]{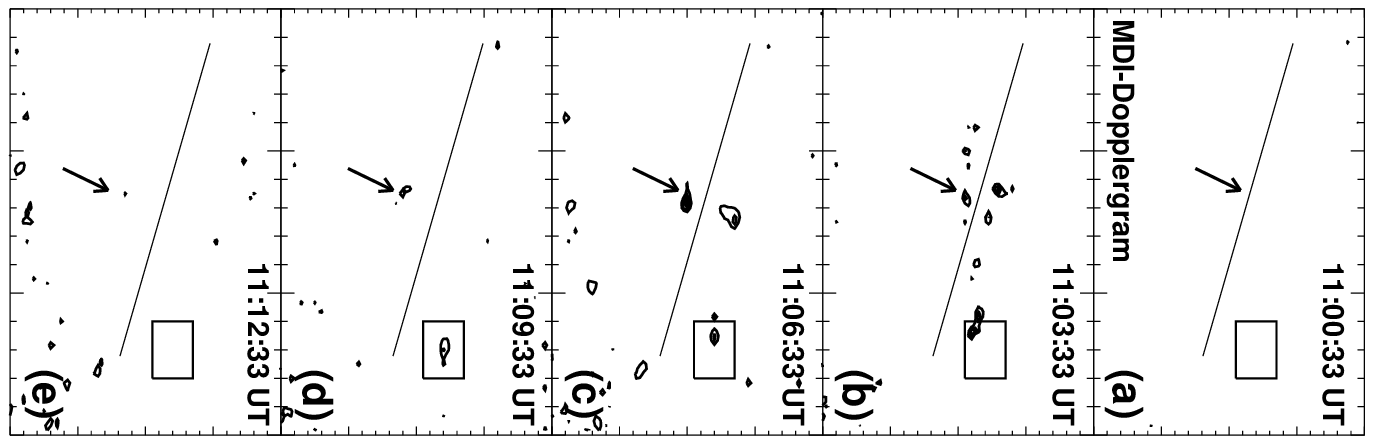}\vspace{-0.098\textwidth}\\
\includegraphics[width=0.3\textwidth, height=0.6\textheight,angle=-90]{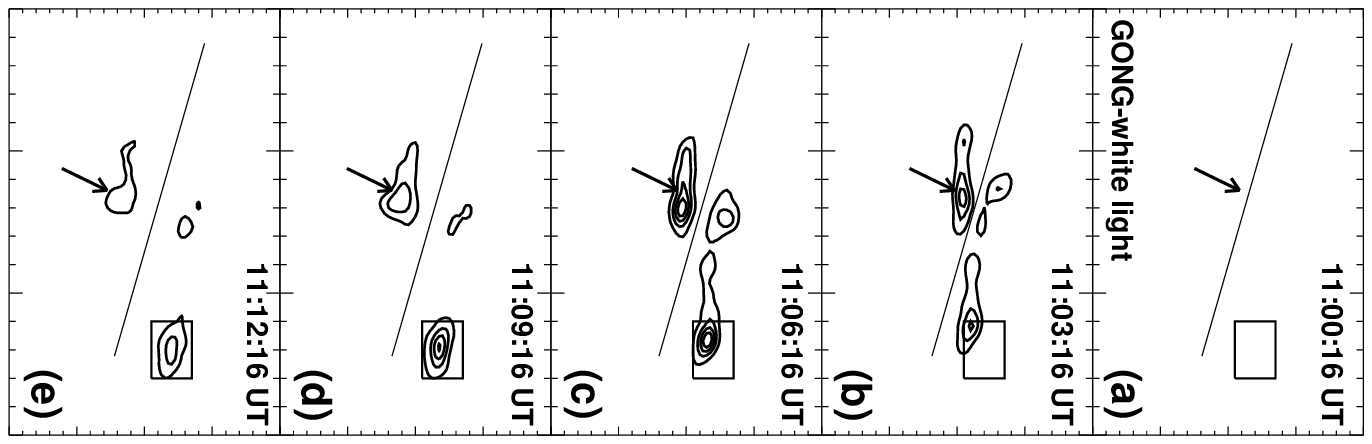}\vspace{-0.098\textwidth}\\
\includegraphics[width=0.3\textwidth, height=0.6\textheight,angle=-90]{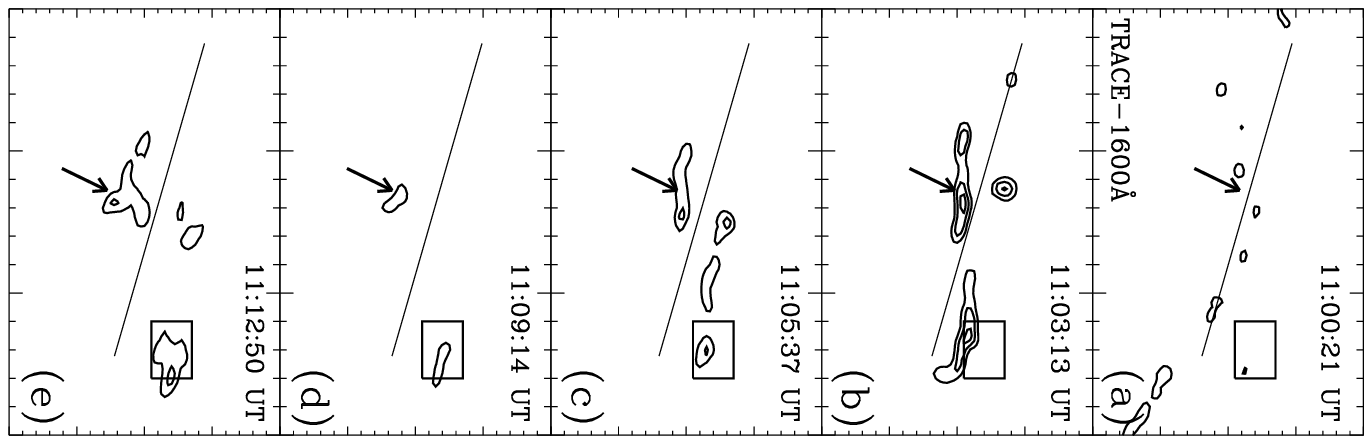}\vspace{-0.098\textwidth}\\
\includegraphics[width=0.3\textwidth, height=0.6\textheight,angle=-90]{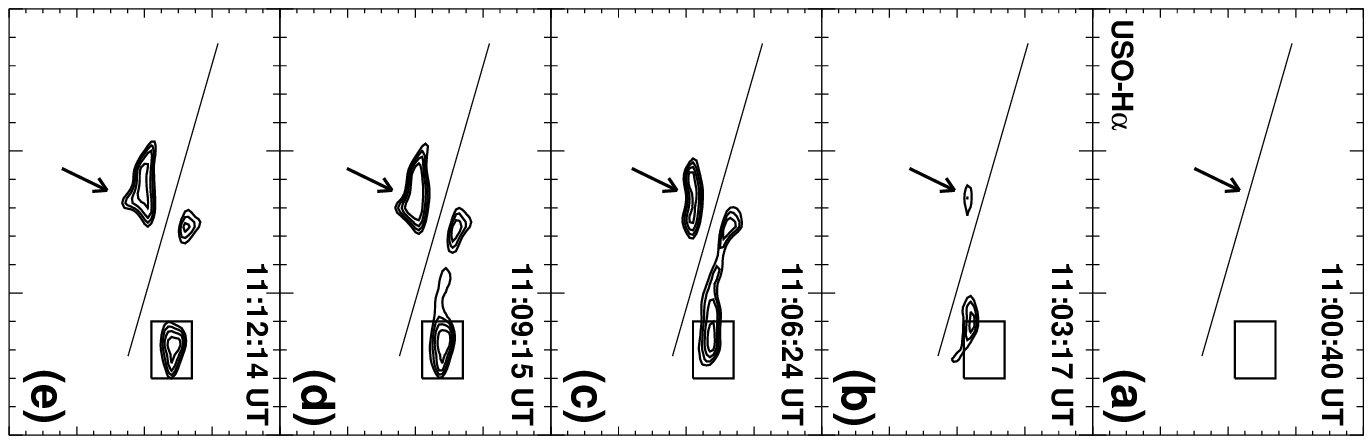}\vspace{-0.098\textwidth}\\
\includegraphics[width=0.3\textwidth, height=0.6\textheight,angle=-90]{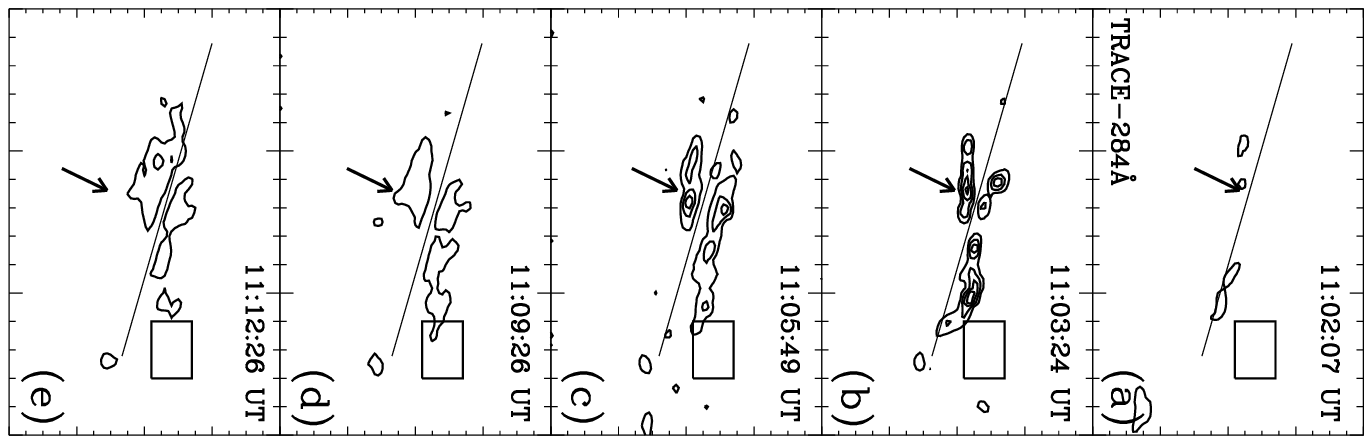}\\
\caption{A mosaic of various observations obtained with time during the impulsive phase of the super-flare of 28 October 2003: MDI magnetograms, MDI Dopplergrams, GONG white-light, TRACE UV 1600\AA, USO H$\alpha$ and TRACE UV 284\AA~(from left to right columns). Appropriate contours (at 20, 40, 60 and 80$\%$ levels of the maximum values in the enhanced images) are shown corresponding to the magnetic/Doppler transients and flare kernels. Arrows and rectangular boxes mark locations of the observed transients.}
   	\label{fig:contours28}
\end{figure}  
\subsection{Motion of the Magnetic/Doppler Transients and  Flare kernels}
    \label{S-motion28}

On a visual inspection of magnetograms/Dopplergrams and filtergrams of the active region taken at various wavelengths, the magnetic transients appeared to be spatially correlated with flare kernels. For establishing this association further, we enhanced the transients using maximum entropy method (MEM) \cite{Skilling84,Narayan86}. It is a deconvolution algorithm which operates by minimizing a smoothness function (``entropy'') in an image. Before applying this tool for each type of observation, we subtracted the square root of a reference image from the square root of corresponding images obtained at different instants of time. Figure~\ref{fig:contours28} (a--e) shows contours of suitably enhanced features observed during the impulsive phase of the flare in different sets of images, \ie, MDI magnetograms and Dopplergrams, GONG white-light, TRACE UV 1600\AA, USO H$\alpha$ and TRACE UV 284\AA. These contours display the spatial and temporal evolution of magnetic/Doppler transients and flare kernels observed at various atmospheric layers with increasing height from the photosphere to the transition region. The solid line in the figure represents an arbitrary reference line, RS, drawn along the neutral line from where distances of various features were measured ({\cf, } Figure~\ref{fig:imbd28}). 
\par It is evident that magnetic/Doppler transients and flare kernels separated away from the neutral line with time during the impulsive phase of the flare (rows a--e, in each column). This follows the usual behavior of two-ribbon flares. We also note that the evolution of magnetic transients (column 1) is spatially and temporally well correlated with the flare kernels observed in different layers (columns 3--6). Locations of observed transients are marked by arrows and boxes in each frame. Along with magnetic transients, comparatively fainter Doppler transients are also discernible during this period (column 2). These are similar to the Doppler-ribbons reported by \inlinecite{Venkatakrishnan08}. Magnetic and Doppler transients are observed to be particularly well correlated with WLF kernels observed during the impulsive phase of the flare.

\begin{figure}
	\centering
		\includegraphics[width=0.98\textwidth]{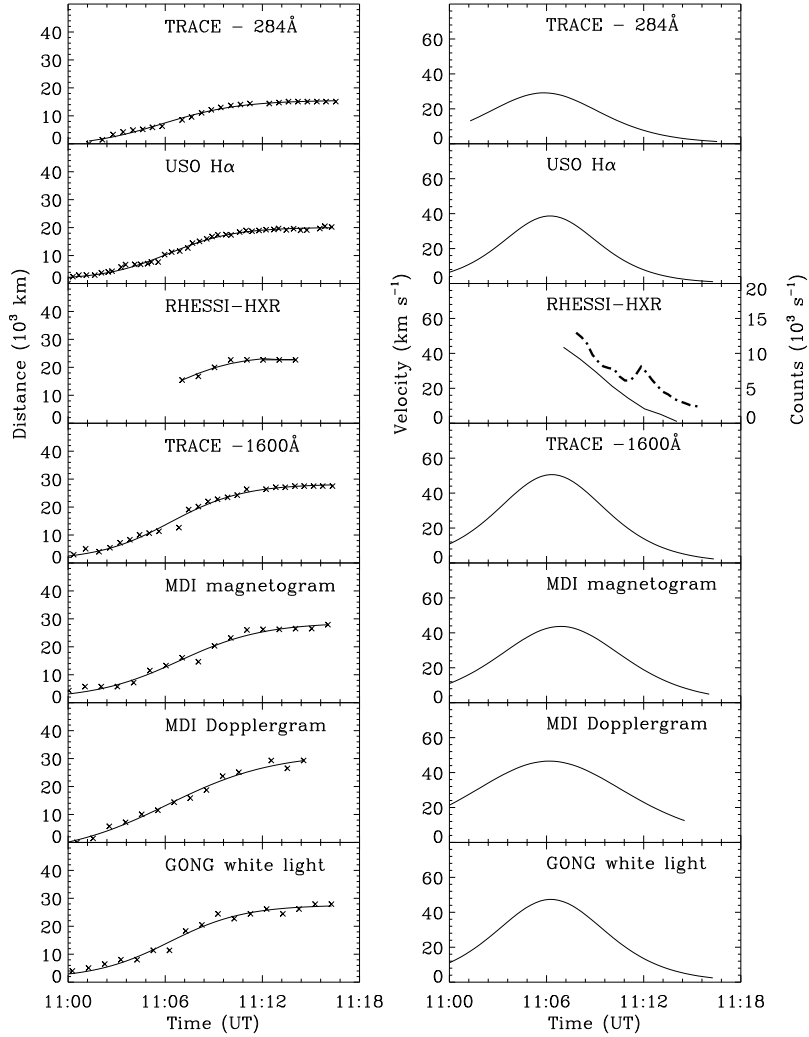}
	\caption{Left panel: Distances of  flare kernels and magnetic/Doppler transients from a reference line during 11:00-11:15 UT using different observations: TRACE UV 284\AA, H$\alpha$, \mbox{RHESSI HXR}, TRACE UV 1600\AA, magnetogram, Dopplergram and white-light (from top to bottom rows). Measured distances are marked by ``$\times$'', while the solid profiles represent the Boltzmann sigmoid fitting through these points. Right panel: Separation velocities derived for various features from the fitted distances corresponding to the left panel. The dash-dotted line in RHESSI velocity panel shows the HXR light curve.}
	\label{fig:distvel28}
\end{figure}%

\par To quantitatively examine motion of the observed features, we carried out the following procedure. Positions of transients and flare kernels were measured from a reference point E  (Figure~\ref{fig:imbd28}, left panel) using an automatic algorithm implemented in IDL (Interactive Data Language). The maximum value was followed along EA, \ie, the direction of motion of features in the enhanced images obtained from MEM. Distance between the point of maximum value within a given feature (say E$^\prime$: $x_1, y_1$) and the reference point E ($x_2, y_2$) was calculated in pixels using $d = [(x_2-x_1)^2 +(y_2-y_1)^2]^{1/2}$. It was then converted into kilometers using the pixel resolution of the corresponding instrument (Table~\ref{T-distvel28}). 

\sloppy{
\begin{table}[ht]
\caption{Maximum distances, velocities, and time at the maximum velocities of transients.}
\label{T-distvel28}
\begin{tabular}{lllllll}
   \hline
S.N. & Data & Pixel res.& Layer              & Max. dist.& Max. vel.& Time of max.\\  
     &       & (arcsec)&                      &   (Mm)     & (km s$^{-1}$)& vel. (UT)\\      
   \hline
1  & TRACE UV 284\AA    & 0.5   &  TR$^{1}$  &  15.4$\pm$0.3    & 29$\pm$4    & 11:05:40 \\
  
2  & USO H$\alpha$   &  0.4  &  Chr$^{2}$ &  20.0$\pm$0.3    & 38$\pm$5    & 11:06:20 \\
  
3  & RHESSI HXR           & 4.0   & Chr$^{2}$   &  23.1$\pm$0.8    &  43$\pm$9      & 11:07:03 \\

4  & TRACE UV 1600\AA   & 0.5   &  TM$^{3}$  &  28.0$\pm$0.3    & 50$\pm$7    & 11:06:13 \\

5  & GONG WL         & 2.5   &  Pho$^{4}$ &  27.3$\pm$0.5    & 45$\pm$6    & 11:06:10 \\
 
6  & MDI Dopplergram      & 2.0   &  Pho$^{4}$ &  29.3$\pm$0.5    & 46$\pm$5    & 11:06:10 \\
 
7  & MDI magnetogram     & 2.0   &  Pho$^{4}$ &  27.9$\pm$0.4    & 43$\pm$4    & 11:06:49\\
 
   \hline
\end{tabular}

\footnotemark[1]{TR: Transition Region,}
\footnotemark[2]{Chr: Chromosphere,}\\
\footnotemark[3]{TM: Temperature Minimum,}
\footnotemark[4]{Pho: Photosphere}
\end{table}
}

It is evident from Figure~\ref{fig:distvel28} that magnetic/Doppler transients and flare kernels separated away from the neutral line during the impulsive phase of the X17/4B flare, \ie, 11:00-11:15 UT. Their distances from the reference point E, along AB, increased rapidly during the initial phase, \ie, 11:00-11:10 UT, as seen in each row (corresponding to different observations) in the left column of the figure. Thereafter, the rate of increase of separation slowed down. Table~\ref{T-distvel28} shows an interesting trend in the maximum separation attained for the flare kernels. It was shortest, \ie, $\approx15$ Mm at the transition layer observed in UV 284\AA~(TRACE), and moving toward the lower layers, it increased to $\approx20$ Mm at the chromosphere (USO H$\alpha$), $\approx28$ Mm at the temperature minimum region and the photosphere (\cf, TRACE UV 1600\AA, and GONG and MDI photospheric data). This trend is consistent with the classical flare model of an expanding loop structure with footpoints anchored in the lower atmosphere. 

\begin{figure}[ht]
	\centering
\includegraphics[width=0.92\textwidth]{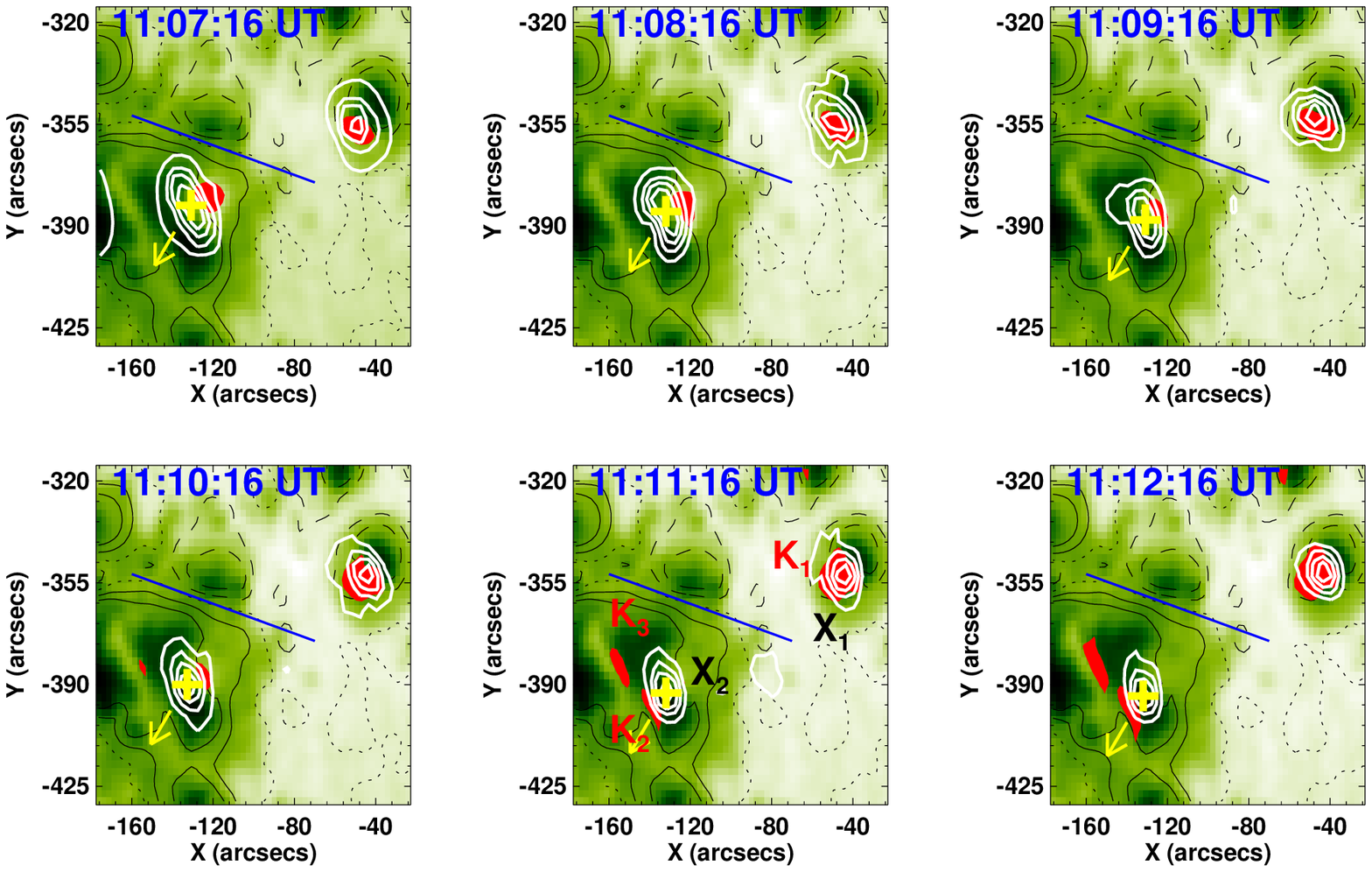}
\includegraphics[width=0.92\textwidth]{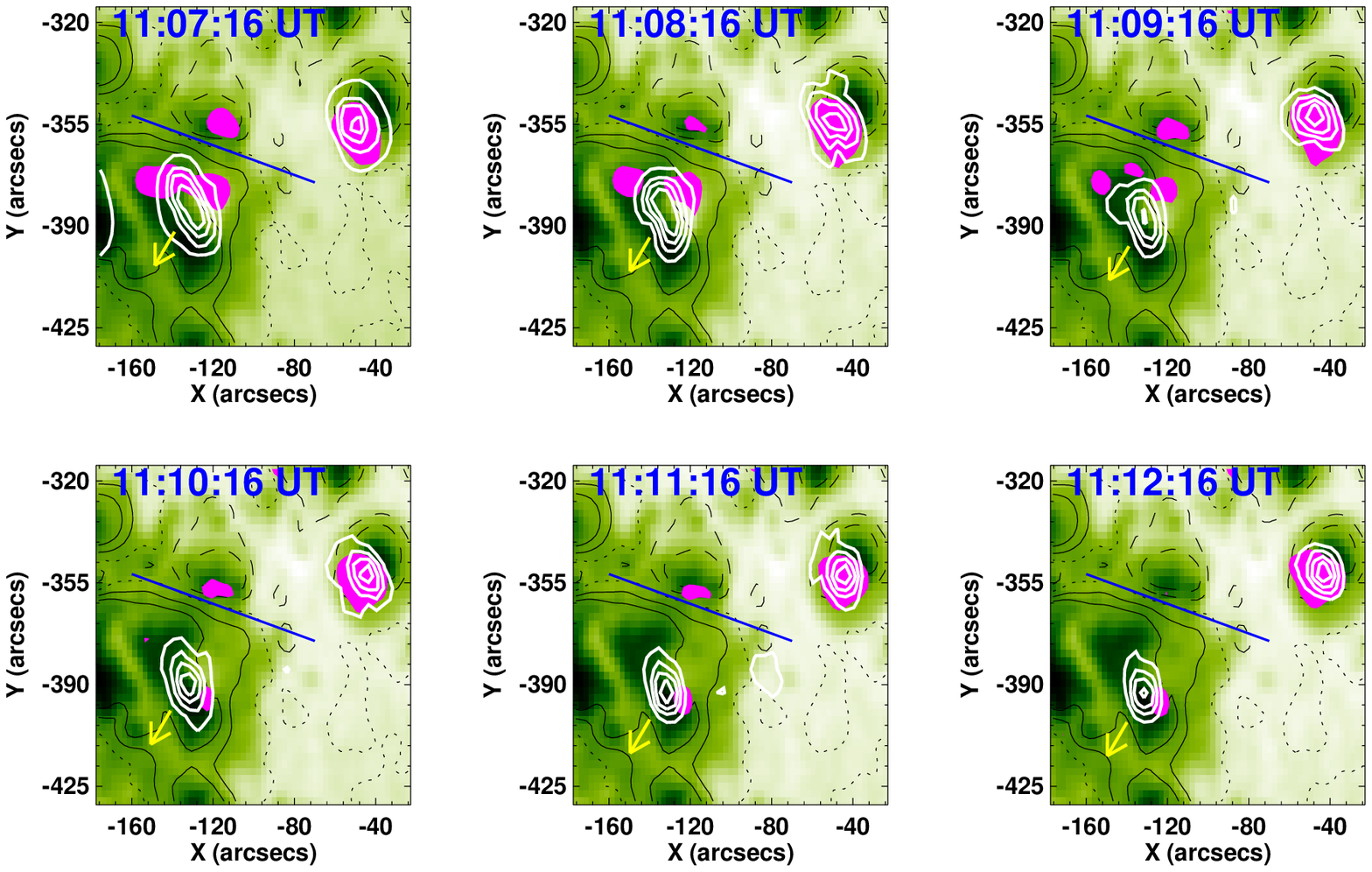}
	\caption{Time sequence of white-light intensity maps (background green half-tone  image) during the impulsive phase of the X17/4B flare - 11:07:16-11:12:16 UT on 28 October 2003. Superimposed dashed (solid) contours represent negative (positive) flux  at $\pm$ 800, $\pm$ 1500 gauss levels, while the dotted contours mark the magnetic neutral lines.  RHESSI HXR fluxes (white contours) in the energy range 100--200 keV are overlaid at the levels 20, 40, 60, 80$\%$  of the maximum value. Red patches mark the enhanced WLF kernels (top rows 1-2) and magenta patches represent the enhanced magnetic transients (bottom rows 3-4).}  
\label{fig:hessi28}
\end{figure}

The HXR feature also showed a separation of $\approx23$ Mm, similar to that observed in the chromospheric H$\alpha$. This is consistent with the formation of HXR footpoint sources in the chromosphere or in the lower corona depending on the energy of penetrating particles. The magnetic and Doppler transients, on the other hand, showed a larger separation of $\approx29$ Mm as obtained for the photospheric WLF kernels. HXR light curve is shown as the dash-dotted line in HXR velocity panel of Figure~\ref{fig:distvel28}. It is evident that the separation velocity of HXR footpoint $\rm{X_2}$ ({\cf,} Figure~\ref{fig:hessi28}) and the overall HXR flux decreased with time during the post-flare phase, except that there is a small bump in the HXR flux at around 11:12 UT. On examining the HXR images we found that the bump corresponds to an enhancement of HXR flux near the footpoint $\rm{X_1}$. It may be noted that the major contribution to the HXR light curve arises from the integrated flux over both the footpoints $\rm{X_1}$ and $\rm{X_2}$ while the separation velocity plotted here pertains only to the footpoint $\rm{X_2}$.         

\par We derived the relative velocities of various features moving along the direction AB, away from the reference line RS, by fitting a Boltzmann sigmoid (best fitting) to the observed positions. This is given by
\[
d(t)=A_0+\frac{A_1-A_0}{1+e^{(t-A_2)/A_3}}
\]
where, $A_0, A_1, A_2$ and $A_3$ are fitting parameters; $A_0$ - top (\ie, maximum of $d(t)$), $A_1$ - bottom (\ie, minimum of $d(t)$),  $A_2$ - the time at which distance is halfway between bottom and top, and $A_3$ - the steepness of the curve, with a larger value denoting a shallower curve.

\par Separation velocity $v(t)$ is then obtained by taking time derivative of the fitted distance function $d(t)$. Numerical differentiation was carried out by three point Lagrangian interpolation. Figure~\ref{fig:distvel28} shows the fitted distances (solid curves) passing through the measured distances marked by ``$\times$'' symbols (left panel), and the derived velocities (right panel). Velocities of flare kernels and magnetic/Doppler transients reached a peak (or maximum) at around 11:06 UT. It is also clear from this figure and Table~\ref{T-distvel28} that the separation velocity of flare kernels was larger, \ie, 45-50 km s$^{-1}$ at lower atmosphere (temperature minimum region and photosphere) and it decreased with height, \ie, 38 km s$^{-1}$ at the chromosphere and 29 km s$^{-1}$ at the transition region. 

\begin{figure}[ht]
	\centering
\includegraphics[width=0.6\textwidth]{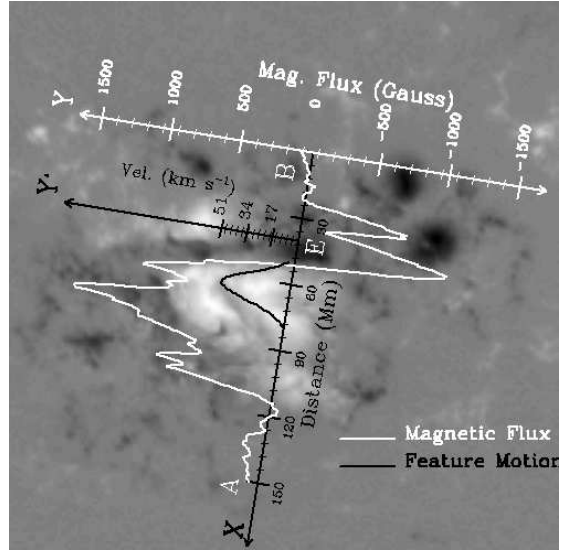}
	\caption{Separation velocity profile of the moving ``magnetic'' feature (dark) measured along the reference line BA overlaid on the MDI magnetogram (background half-tone image). The magnetic flux variation along the trajectory of motion of the feature is drawn as white profile.}
	\label{fig:flux_BA}
\end{figure} 

\par It is known that a systematic increase in the separation of H$\alpha$ flare kernels is the chromospheric signature of magnetic reconnection process occurring progressively at higher levels in the corona. Rather small separation velocities in the range of $\approx$ 3-10 km s$^{-1}$ have been reported for two-ribbon flares associated with filament eruptions as compared to the velocities in the range of 30-50 km s$^{-1}$ obtained for the flares in NOAA 10486 (Table~\ref{T-distvel28}). However, large separation velocities in the range 20-100 km s$^{-1}$ have also been reported during some recent two-ribbon flares occurring in complex active regions \cite{Qiu02}. 

As the flare-ribbons approach regions of strong magnetic fields, their speeds are known to decrease. This is also observed for the flare kernels and moving magnetic transients during the flare of 28 October 2003. Figure~\ref{fig:flux_BA} shows the separation velocity profile (dark curve) of the moving magnetic feature from the neutral line along the direction BA. It is clear that the separation velocity first increased with magnetic flux (white curve), attained a peak value, then decreased and vanished at the location where magnetic flux attained a maximum value. This is consistent with the general behavior of flare-ribbons moving toward the region of strong magnetic fields.

\subsection{RHESSI HXR Sources and the White-light Flare of 28 October 2003}
    \label{HXR-motion28}

Figure~\ref{fig:hessi28} shows temporal and spatial evolution of enhanced WLF kernels (red patches) observed in GONG intensity maps (green half-tone images) obtained during the impulsive phase of the X17/4B flare of 28 October 2003 (top rows 1-2). Similarly, evolution of moving magnetic transients (magenta patches) is shown over the same background images in the bottom rows (3-4). White contours represent RHESSI HXR flux in the energy range 100-200 keV associated with the flare during its impulsive phase, \ie, 11:07:03-11:12:03 UT, at 20, 40, 60, 80$\%$ levels of the maximum HXR flux. The central maximum of the HXR source (marked by ``+'' sign in yellow color) was found to be moving in the direction of arrow with separation velocity of 43 km s$^{-1}$ from the reference line, \ie, the solid blue line drawn along the neutral line. Separation velocity of the HXR feature was similar to that of the moving magnetic/Doppler transient and WLF kernels (\cf, Table 1), however, their spatial positions differed some what. 

HXR features were earlier reported to form during the impulsive phase of an X5.6 flare of 6 April 2001 in strong magnetic field regions \cite{Qiu03}. Figure~\ref{fig:contours28} shows that the moving magnetic transient, where polarity sign reversal occurred, first appeared around 11:03 UT close to the neutral line, \ie, at a weak magnetic field location. Thereafter, it moved toward the direction of stronger magnetic field of the following (positive) polarity umbra. Unfortunately, we do not have the entire temporal sequence in RHESSI HXR due to the non-availability of data before 11:07:16 UT. Therefore, it is not clear whether the HXR source had also formed near the neutral line in the initial phases.  

\inlinecite{Qiu03} noted that locations of sign-reversal ``anomaly'' coincided well with HXR footpoints. They suggested the observed anomalous magnetic features to be related to electron precipitation into the lower atmosphere but not with all WLF kernels. They gave the energetics for the reversal of line-profiles in the cooler, strong magnetic field umbral regions as compared to that in the hotter, weak magnetic field locations. In agreement with \inlinecite{Qiu03}, we observed that HXR footpoint $\rm{X_1}$ formed in this case also at strong magnetic field location of the leading umbra. This HXR source was nearly stationary and matched well with both the WLF kernel $\rm{K_1}$ (rows 1-2) and the magnetic transient feature (rows 3-4). On the other hand, HXR source $\rm{X_2}$ did not match well with the WLF kernel. We observed that the WLF kernel $\rm{K_2}$ first appeared at one edge of $\rm{X_2}$ contours (11:07:16 UT), then decayed and appeared at the opposite edge (11:11:16 UT). Around the same time, WLF kernel $\rm{K_3}$ appeared which was not associated with any HXR feature. The moving magnetic feature at $\rm{X_2}$ appear to be better associated with HXR than WLF kernels. Thus, association of HXR source with anomalous magnetic polarity feature, as reported by \inlinecite{Qiu03}, does appear to hold in this case.
\begin{figure}[ht]
	\centering
\includegraphics[width=1.0\textwidth]{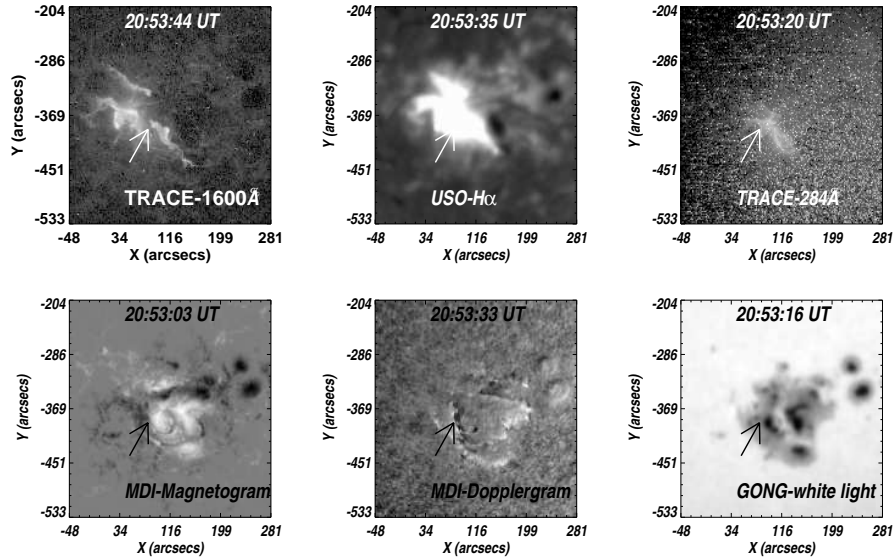}
	\caption{TRACE UV 1600\AA, MLSO H$\alpha$ and TRACE UV 284\AA~(top row, from left to right columns) and the MDI magnetogram, Dopplergram and GONG white-light image (bottom row, from left to right columns) obtained during the peak phase of the white-light X10/2B flare of 29 October 2003. Arrow shows the location of the ``moving'' magnetic transient. }
	\label{fig:immos29}
\end{figure} 
\section{The X10/2B Flare of 29 October 2003/20:49 UT}
  \label{S-flare29}
This white-light flare, classified as X10/2B, occurred in NOAA 10486 on 29 October 2003 when the active region was located close to the disk center at S15W02. The flare began at 20:37 UT, reached peak phase  at 20:49 UT and decayed at 21:01 UT. Figure~\ref{fig:immos29} shows a mosaic of flare images as observed in the TRACE UV 1600\AA, H$\alpha$ and TRACE UV 284\AA~(top row). The photospheric magnetogram, Dopplergram and white-light images are shown in the bottom row. As detected during the white-light flare of 28 October 2003, we found moving transient features around the impulsive phase of this energetic event also. However, this flare occurred in another part of the active region away from the event of 28 October 2003. Locations of magnetic transients and WLF kernels are marked by arrow in each frame of Figure~\ref{fig:immos29}. The characteristics of MFs and flare kernels are described in the following.

\begin{figure}[ht] 
	\centering
		\includegraphics[width=1.0\textwidth]{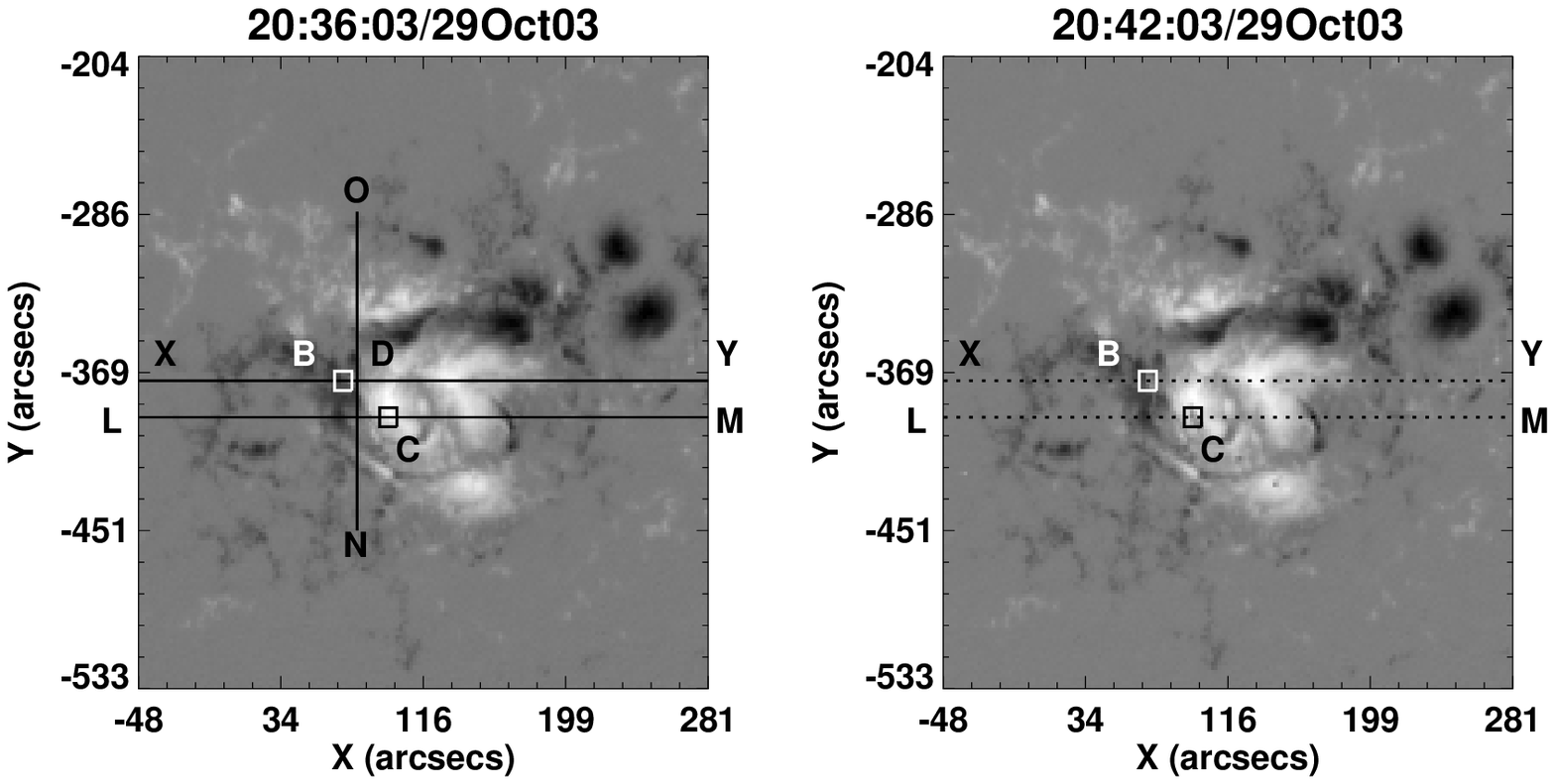}
	  \caption{MDI magnetograms of NOAA 10486 taken around the pre- and peak phases of the 29 October 2003 flare at 20:36:03 UT (left panel) and 20:42:03 UT (right panel), respectively. Magnetic flux along the lines XY and LM is drawn in Figure~\ref{fig:mflux29}. Sign reversals in flux polarity were detected at ``B'' and ``C'' across the neutral line. Point ``D'' on the vertical line NO represents the reference from which distances of flare kernels were measured along line DX.}
	\label{fig:imbdf29}
\end{figure}
\begin{figure}[ht] 
	\centering
		\includegraphics[width=0.9\textwidth]{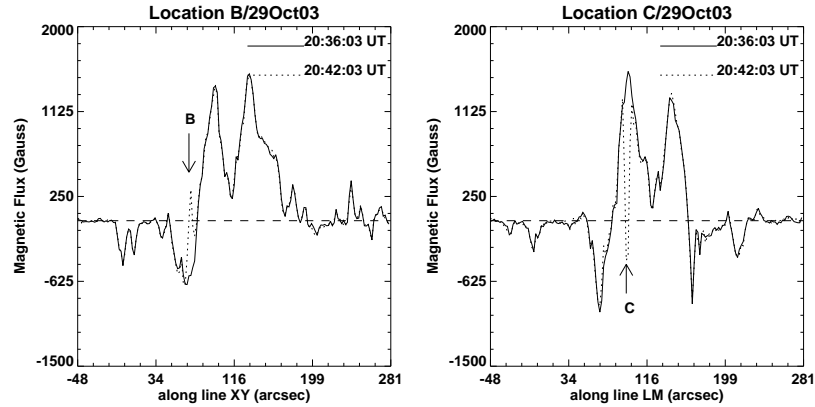}
		\caption{Magnetic flux along the lines XY and LM (as shown in Figure~\ref{fig:imbdf29}). Solid and dotted profiles represent magnetic flux obtained  at 20:36:03UT and 20:42:03 UT, respectively. Polarity sign reversals are evident at the locations B and C marked in Figure~\ref{fig:imbdf29} during the peak phase of the white-light flare at 20:42:03 UT, whereas the flux remained nearly unchanged else where along XY and LM.}
	\label{fig:mflux29}
	\end{figure}
\begin{figure}[ht]
   \centering
\includegraphics[width=0.3\textwidth, height=0.6\textheight,angle=-90]{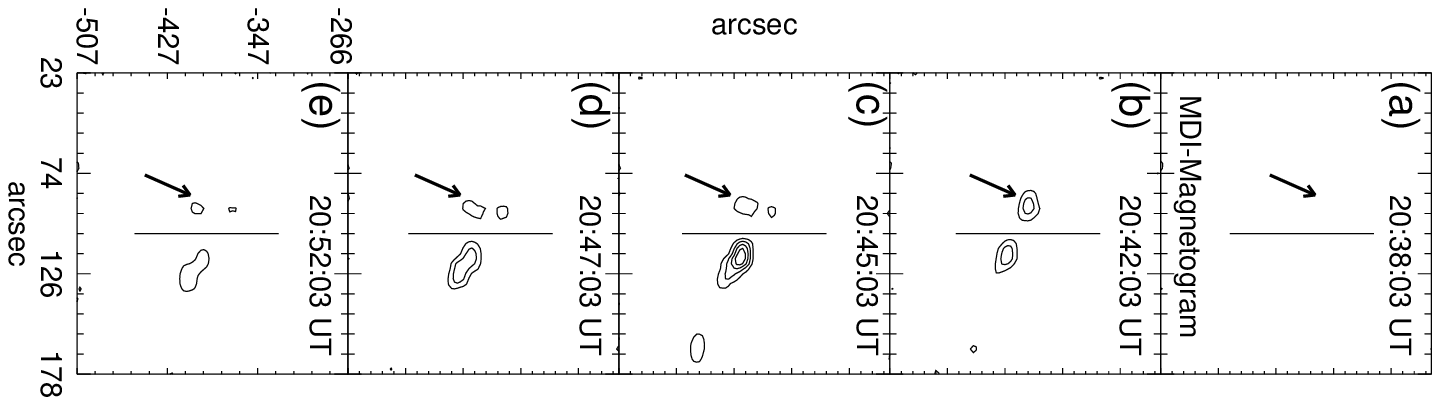}\vspace{-0.075\textheight}\\
\includegraphics[width=0.3\textwidth, height=0.6\textheight,angle=-90]{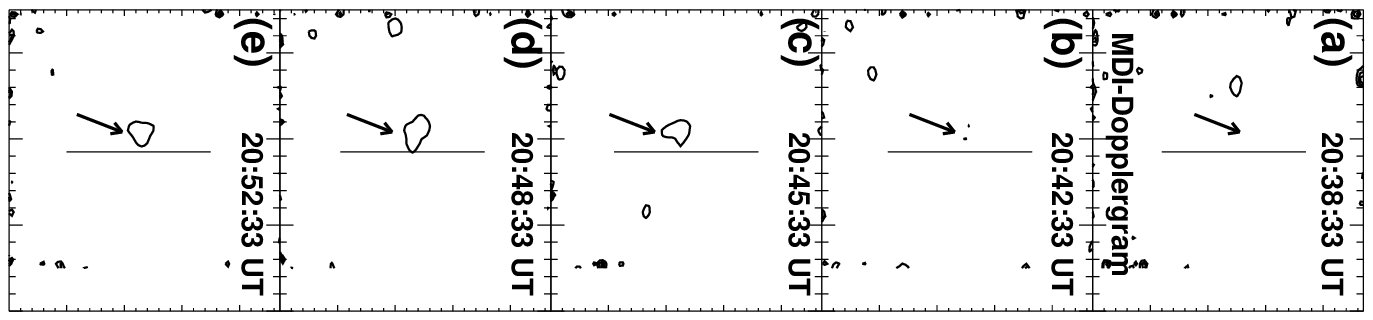}\vspace{-0.122\textwidth}\\
\includegraphics[width=0.3\textwidth, height=0.6\textheight,angle=-90]{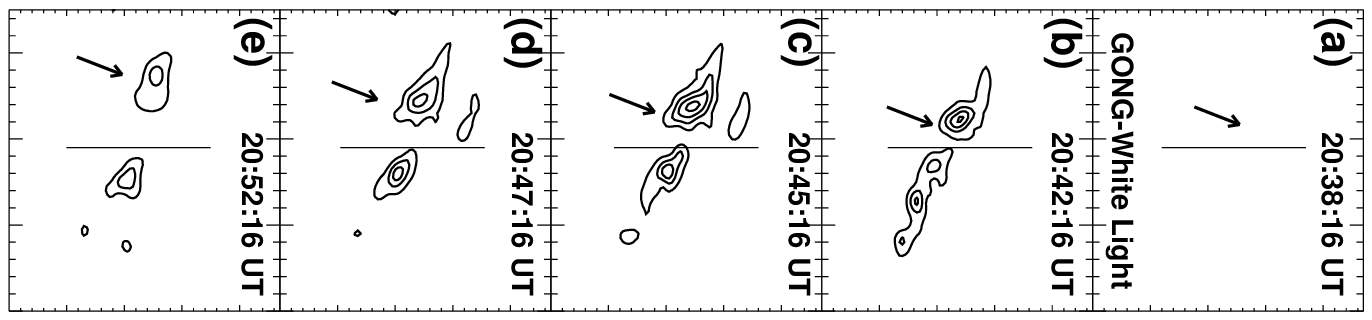}\vspace{-0.122\textwidth}\\
\includegraphics[width=0.3\textwidth, height=0.6\textheight,angle=-90]{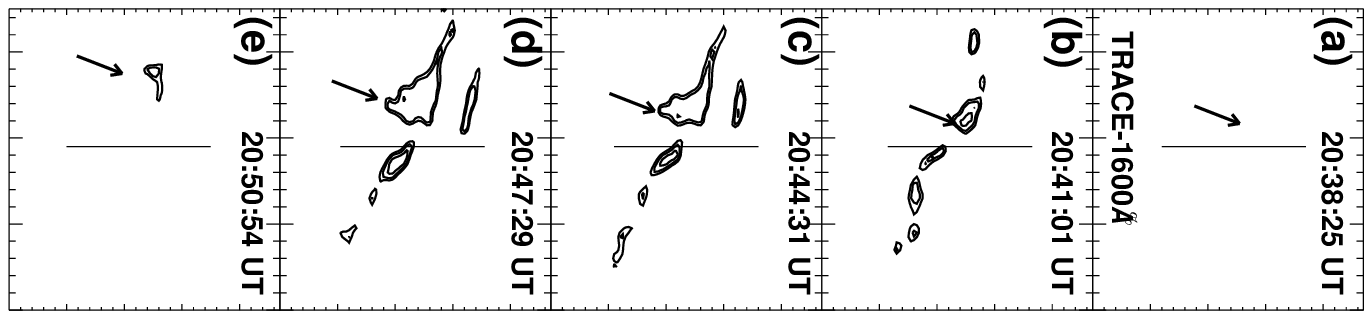}\vspace{-0.122\textwidth}\\
\includegraphics[width=0.3\textwidth, height=0.6\textheight,angle=-90]{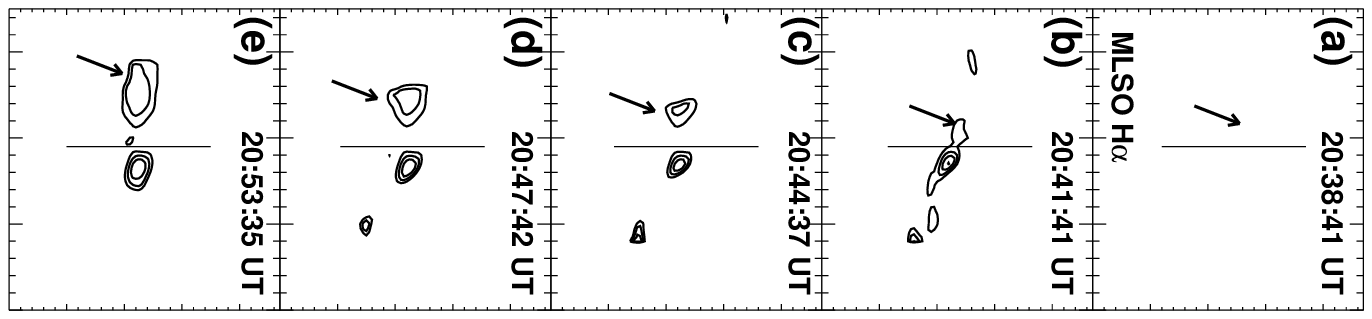}\\
\caption{A mosaic of contours of different observations: magnetograms (MDI), Dopplergrams (MDI), white-light (GONG),  UV 1600\AA~(TRACE) and H$\alpha$ (MLSO) (from left to right columns). Rows correspond to different time instants of the X10/2B flare of 29 October 2003 during its impulsive phase. Arrows have been drawn at the locations of MF and at corresponding positions in the flare kernels. The solid line is the reference direction with respect to which MF and flare kernel motions were measured.}
	\label{fig:magmos29}
\end{figure}
\begin{figure}
	\centering
		\includegraphics[width=0.6\textwidth]{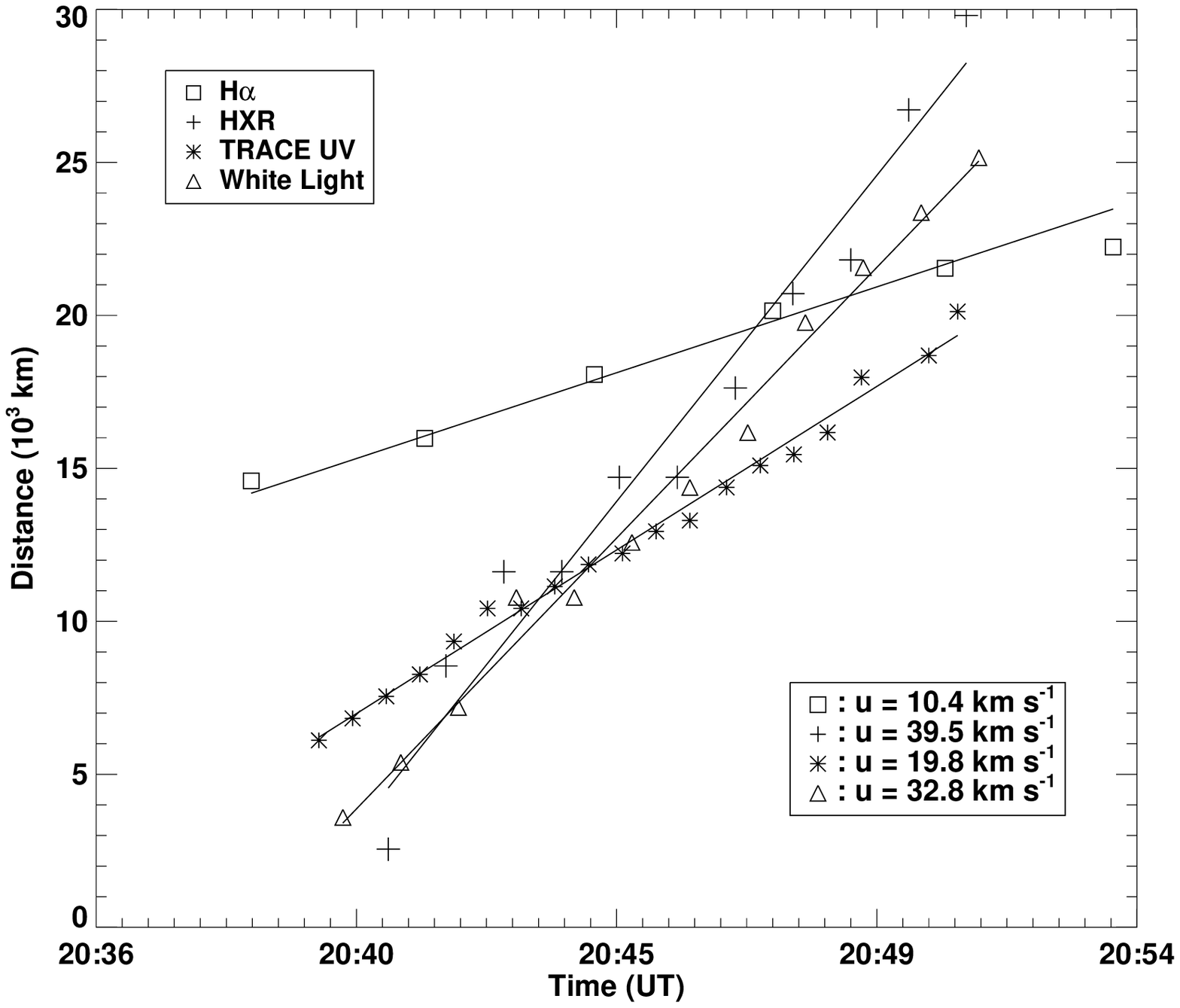}
	\caption{Distances of flare kernels from the reference line NO during 20:38:40-20:53:35 UT for different observations: H$\alpha$ (MLSO), HXR (RHESSI), UV 1600\AA~(TRACE) and white-light (GONG). Measured distances are marked by symbols $\Box$, $+$, $*$ and $\bigtriangleup$ for H$\alpha$, HXR,  UV 1600\AA, and WL, respectively, while solid lines represent the straight-line fitting through the corresponding data points.}
	\label{fig:distvel29}
\end{figure}

 \subsection{Magnetic Field Transients and the Sign Reversal }
 \label{S-inversion29}
A movie of MDI magnetograms was used to study the evolution of magnetic flux and detect any magnetic transient features during the impulsive phase of this flare also (\cf, Figure~\ref{fig:imbdf29}). We identified locations B and C where polarity sign reversals occurred at 20:42:03 UT by examining magnetic flux along a horizontal raster moving from the bottom to the top of magnetograms selected during the pre-peak and peak phases of the flare. Magnetic flux along XY and LM is plotted in Figure~\ref{fig:mflux29}. Solid and dotted lines in the figure correspond to the times of pre-peak phase (20:36:03 UT) and the peak phase (20:42:03 UT), respectively. It is observed that magnetic flux changed sign around B and C, while it remained nearly unchanged else where along the lines XY and LM. 
\subsection{Motion of Magnetic/Doppler Transients and Flare Kernels} 
\label{S-motion29}
A careful examination of MDI magnetogram and Dopplergram movies of the 29 October 2003 event revealed moving magnetic and Doppler features along with WLF kernels during the impulsive phase. Spatial and temporal evolution of MFs and flare kernels observed during this event are illustrated in Figure~\ref{fig:magmos29}. Contour levels were derived from the same technique (MEM) as used for the previous event of 28 October 2003. Different sets of images, \ie, photospheric magnetograms (MDI), Dopplergrams (MDI), white-light (GONG), chromospheric H$\alpha$~(MLSO) and temperature minimum UV 1600\AA~(TRACE) are shown from the left to the right columns.

The MFs appeared to be correlated with flare kernels observed in white-light and other wavelengths corresponding to various layers of the solar atmosphere. Temporal evolution of MFs and flare kernels is shown from the top (a) to the bottom (e) rows in each column. A solid line drawn in each frame marks the reference direction perpendicular to which MFs and flare kernels moved, and the arrow shows the location of the MF in each set of the observations.  

\sloppy{
\begin{table}[ht]
\caption{Maximum distances and velocities of the flare kernels.}
\label{T-distvel29}
\begin{tabular}{lllllll}
   \hline
S.N. & Data & Pixel res.& Layer     & Max. dist.& Max. vel.\\  
     &       & (arcsec)&          &   (Mm)     & (km s$^{-1}$)\\      
   \hline
 
1  & USO H$\alpha$  &  0.4  &  Chr$^{2}$ &  23.5$\pm$0.3    & 10$\pm$5     \\

2  & RHESSI HXR  & 4.0    & Chr$^{2}$  & 28.3$\pm$0.9    &  40$\pm$9       \\
  
3  & TRACE UV 1600\AA& 0.5  &  TM$^{3}$  &  19.3$\pm$0.3    & 20$\pm$7    \\

4  & GONG WL& 2.5   &  Pho$^{4}$ &  25.0$\pm$0.5    & 33$\pm$6     \\
 
   \hline
\end{tabular}

\footnotemark[1]{TR: Transition Region,}
\footnotemark[2]{Chr: Chromosphere,}\\
\footnotemark[3]{TM: Temperature Minimum,}
\footnotemark[4]{Pho: Photosphere}
\end{table}
}    

\par Distances of flare kernels from the reference line NO along the line DX are determined by following the position of the maximum value within a feature of interest.  Separations of various flare kernels are shown in Figure~\ref{fig:magmos29}. In order to find separation speeds of flare kernels, we fitted straight lines (best fitting) through the measured distances. Slopes of these fitted lines provided the respective velocities  (Figure~\ref{fig:distvel29}). Maximum distances and velocities are listed in  Table~\ref{T-distvel29}. Velocities of flare kernels showed similar trend as in the earlier case of the flare of 28 October 2003; it generally increased from the upper to the lower atmospheric layers. However, separation velocities obtained for this flare were found to be much smaller as compared to the X17/4B flare. 

Another difference was that the magnetic and Doppler transients remained nearly stationary during this flare. The profiles for distance and separation velocity were also different from that of the event of 28 October 2003, where we could do a Boltzmann Sigmoid fitting as against the straight line fitting for this flare. It is unlikely that the separation velocity of flare kernels would remain constant throughout the flare duration as appears to be the case in this event. It is to note that flare kernels became fainter in intensity or got fragmented as they moved away from the reference line, and were generally difficult to identify and follow after 20:51 UT. For H$\alpha$ kernel, however, there is an indication that the rate of change of separation decreased after this time.

\begin{figure}[ht]
	\centering
\includegraphics[width=0.92\textwidth]{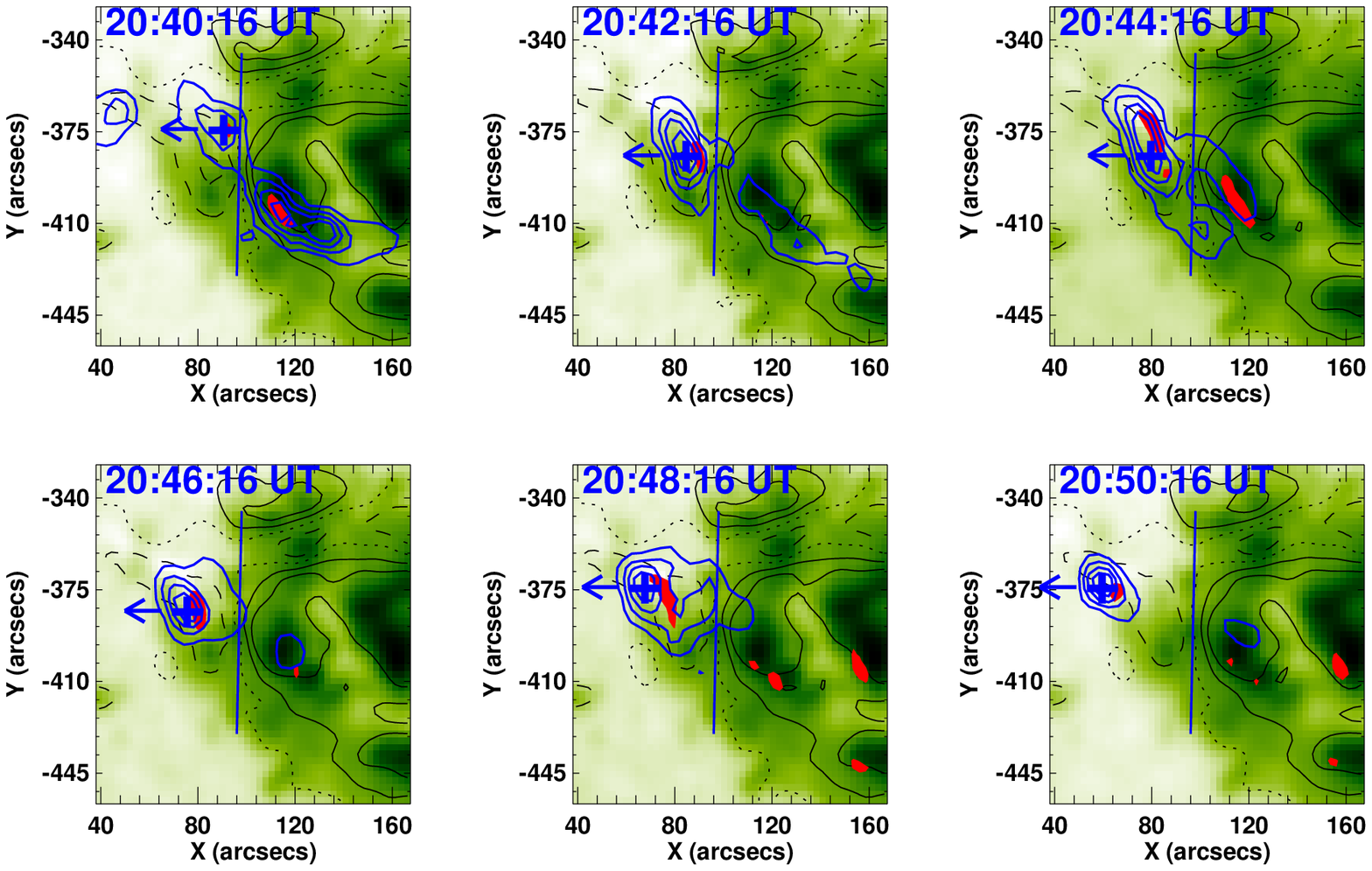}
\includegraphics[width=0.92\textwidth]{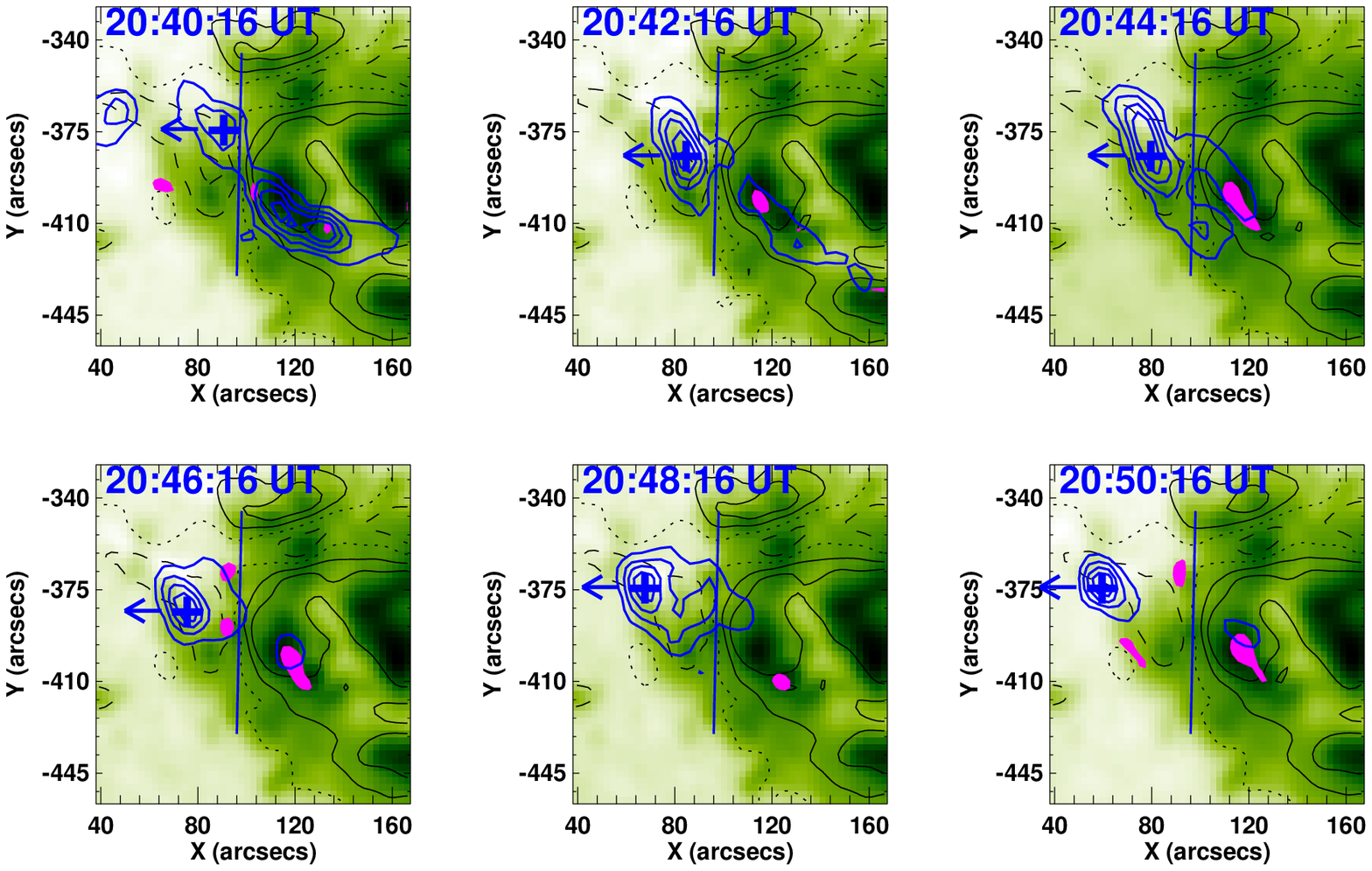}
	\caption{Time sequence of white-light intensity maps (background half-tone green image) during the impulsive phase of the X10/2B flare - 20:40:16-20:50:16 UT on 29 October 2003. Superimposed dashed (solid) contours represent negative (positive) flux levels at $\pm$ 800, $\pm$ 1500 gauss, and dotted contours mark the magnetic neutral lines.  RHESSI HXR fluxes in the energy range 50-100 keV are overlaid  at 20, 40, 60, 80$\%$ levels of the maximum value (blue contours). Red patches (top rows 1-2) show the positions of WLF kernels, while magenta patches (bottom rows 3-4) mark the magnetic transients.}  
	
	\label{fig:hessi29}
\end{figure}    
\subsection{RHESSI HXR Sources and the White-light Flare of 29 October 2003}
    \label{HXR-motion29}
Figure~\ref{fig:hessi29} shows temporal and spatial evolution of the enhanced WLF kernels (red patches) observed in GONG intensity maps (green half-tone images) during the impulsive phase of the X10/2B flare of 29 October 2003 (top rows 1-2). Similarly, evolution of magnetic transients (magenta patches) is shown over the same background images in the bottom rows (3-4).  Blue contours represent RHESSI HXR flux in the energy range 50-100 keV associated with the flare during its impulsive phase, \ie, 20:40:16-20:50:16 UT, at  20, 40, 60, 80$\%$ levels of the maximum flux. The central maximum of the HXR source is marked by ``+'' sign in blue color. A solid blue line is drawn along the neutral line as a reference to measure the positions of various flare kernels observed to be moving in the direction of the arrow.  

\par Corresponding to the white-light flare of 29 October 2003, Xu \etal~\shortcite{Xu04} obtained  1.56 $\mu$m near-infrared (NIR) images of the region $\approx$ 50 km below the photosphere at the opacity minimum layer using the Dunn Solar Telescope at the National Solar Observatory/Sacramento Peak. These images indicated extremely energetic activities associated with this flare as only the most energetic electrons can penetrate this deep in the photosphere. Locations of GONG WLF kernels matched very well with the NIR continuum flare patches. However, not all these WLF kernels were associated with RHESSI HXR sources, as seen in Figure~\ref{fig:hessi29} (top rows 1-2). Also, locations of MFs, where sign-reversal anomaly appeared, did not conform well with these HXR sources except at 20:40:16 UT (first two frames in row 3). Thereafter, the HXR source intensified at the opposite side of the neutral line, where the other part of the two-ribbon WLF kernel was located (top rows 1-2), but no magnetic transient feature (magenta patches in rows 3-4) was present. 

Interestingly, in disagreement with \inlinecite{Qiu03}, we note that the strong HXR source (and the WLF kernel) which intensified at the west side of the neutral line was located in a weak field region. On the other hand, the magnetic transient observed at stronger field positive polarity umbra was generally associated with WLF kernel but no HXR source. This is in contrast with the white-light flare of 28 October 2003, and also contradicts the suggested association of sign reversal with HXR sources. However, it may be noted that \inlinecite{Chen06} found, for a white-light flare of 30 September 2002, that motions of the continuum and HXR sources show similar trend.  
\section{Discussions and Conclusions}
     \label{S-Discussions}
We observed moving transients and sign reversals of magnetic polarity at some locations of NOAA 10486 during the impulsive phases of both the white-light flares of 28 and 29 October 2003. Comparatively fainter Doppler transients were discernible along with magnetic transients during these flares. The magnetic and Doppler transients were observed to be well correlated. Sign reversals in magnetic polarity and Doppler velocity also occurred at the locations of these moving transients.  

\par The two flares were quiet different in their energetics - the 28 October flare was X17/4B, while the 29 October flare was X10/2B. They occurred at topologically different locations within the active region. We have attempted to examine properties and relationship of the observed MFs and flare kernels in various atmospheric layers for the two events. There are some interesting questions related to these observed transients: What kind of physical process caused the moving magnetic and Doppler velocity transients? Is there any relationship between moving transients, HXR sources and flare kernels? 

\par The GONG and MDI instruments use the same spectral line, \ie, Ni~{\sc i} 6768\AA, but different principles of observations (Harvey 2008, private communication). Nevertheless, the observed motions in magnetic, Doppler velocity and intensity fields obtained from both instruments were similar in nature. Sign reversals in magnetic and Doppler velocity fields were detected by both instruments during the impulsive phases of the flares of 28 and 29 October 2003. Some of these transients occurred near the cooler umbral/penumbral sites in the active region NOAA 10486, but there were also several exceptions.  

\par  In the case of the X17/4B flare of 28 October 2003, we found that the two locations of magnetic transients were associated with sign reversals in magnetic polarity (and Doppler velocity). These features differed in their basic characteristics as follows: (i) The location of sign reversal that occurred in strong magnetic field of the leading (negative) umbra was nearly stationary. Both HXR source and WLF kernel were associated with this feature. (ii) The other location of sign reversal observed in the weak field area near the neutral line moved rapidly toward the following (positive) polarity umbra, and was better associated with HXR as compared to the WLF kernels. 

\par On the other hand, magnetic and Doppler transients observed on 29 October 2003 were nearly stationary during the course of the X10/2B flare. The locations of magnetic transients, where sign-reversal anomaly appeared, did not conform well with the HXR sources. Interestingly, we observed a strong HXR source (and WLF kernel) in weak field region without corresponding sign reversal anomaly. On the other hand, there was a WLF kernel but no HXR source associated with the magnetic transient observed in strong field location of the positive polarity umbra. This is in contrast with the HXR-magnetic anomaly relation observed for the white-light flare of 28 October 2003, and also in contradiction with the earlier suggestions of  \inlinecite{Qiu03} about the association of HXR sources with polarity sign reversal phenomenon.

\par We found that magnetic/Doppler transients generally followed the usual behavior of separation of flare kernels observed in two-ribbon flares. Let us first discuss the motion of magnetic transients observed during the flare of 28 October 2003. There was an interesting trend in maximum separation attained for the flare kernels observed in successively lower layers of solar atmosphere: the shortest, \ie, $\approx15$ Mm at the transition layer as observed in UV 284\AA~(TRACE), increasing to $\approx20$ Mm at the chromosphere (USO H$\alpha$) and $\approx28$ Mm in the  temperature minimum region and the photosphere (UV 1600\AA~TRACE, and GONG \& MDI photospheric data). This trend is consistent with the classical flare model of an expanding loop structure with footpoints anchored in the lower atmosphere. The HXR feature also showed separation of  $\approx23$ Mm, \ie, nearly the same as the separation observed in chromospheric H$\alpha$. This is consistent with the formation of footpoint sources in the chromosphere or in the lower corona depending on the energy of particles. Magnetic and Doppler transients, on the other hand, showed a larger separation of $\approx29$ Mm which matched the separation obtained for the photospheric WLF kernels.

\par Furthermore, we found that flare kernels separated faster in the lower atmosphere (the temperature minimum region and the photosphere) with velocity of 45-50 km s$^{-1}$ as compared to that in the upper atmosphere, \ie, 38 km s$^{-1}$ at the chromosphere, and 29 km s$^{-1}$ at the even higher transition region. However, separation velocities for the X10/2B flare of 29 October 2003 were found to be much smaller as compared to the X17/4B flare of the previous day.  

\inlinecite{Ding02}, using an atmospheric model, showed that spectral line inversion occurs near sunspot penumbral area if electron density rises significantly. It is to note that a rise in the electron density was found during the impulsive phase of the flare of 28 October 2003 \cite{Klassen05}, also consistent with the RHESSI HXR observations. Therefore, occurrence of moving transients appears to be related to line profile changes due to the process associated with electron-beams and not due to real changes in photospheric magnetic and Doppler fields. However, no clear conclusions are discernible in the case of the X10/2B flare of 29 October 2003. Also, away from these features, permanent changes in magnetic fluxes from the pre- to post-flare phases have been reported that do not seem to be affected by the WLF or HXR related effects \cite{Ambastha07a}. 

From a study of the flare of 29 September 2002, \inlinecite{Chen05} found that the continuum enhancement at HXR footpoints is not proportional to the electron beam flux, but it depends on the atmospheric conditions. Using H$\alpha$ line profiles they explained that the atmosphere was heated considerably during the impulsive phase, which increased the coronal pressure preventing nonthermal electrons from effectively penetrating into the chromosphere. But during the preflare phase the heating is low that allows an electron beam to easily penetrate into the chromosphere and produce the continuum enhancement via radiative back-warming effect.
\par It is interesting to note that the observed flare kernels and moving transients associated with these events possessed large velocities. This is  expected due to the large magnetic energy stored in the magnetically complex active region which produced these exceptionally energetic super-flares. Large separation velocities of flare-ribbons and HXR footpoints are the signatures of rapid reconnection of successive magnetic field lines at higher levels of corona.

\par Finally, answer to the question of association characteristics of the moving magnetic transients (or magnetic anomaly) still remains ambiguous as a clear and consistent correlation of these features with HXR sources (or WLF kernels) did not emerge from the study of the two white-light flares of NOAA 10486.   
%

\begin{acks}
We are thankful to J. Harvey, NSO, for communicating the detailed description of the principle of GONG observations and its inter-comparison with MDI. 
\par This work utilizes data obtained by the Global Oscillation Network Group (GONG) project, managed by the National Solar Observatory  which is operated by AURA Inc., under a cooperative agreement with the National Science Foundation. The GONG data were acquired by instruments operated by the Big Bear Solar Observatory, High Altitude Observatory, Learmonth Solar Observatory, Udaipur Solar Observatory, Instituto de Astrofisico de Canarias and Cerro Tololo Inter-American Observatory.
\par This work utilises data from the Solar Oscillations Investigation/ Michelson Doppler Imager (SOI/MDI) on the Solar and Heliospheric Observatory (SOHO). SOHO is a project of international cooperation between ESA and NASA. MDI is supported by NASA grants NAG5-8878 and NAG5-10483 to Stanford University.
\par We would like to express thanks for the H$\alpha$ data to Mauna Loa Solar Observatory (MLSO). MLSO is operated by the High Altitude Observatory (HAO), a division of the National Center for Atmospheric Research (NCAR) and funded by the National Science Foundation (NSF).
\par This work also utlises data obtained by Ramaty High-Energy Solar Spectroscopic Imager.
\end{acks}


\end{article}
\end{document}